\documentclass[secnumarabic, twocolumn, graphics,floatfix, superscriptaddress,tightenlines, aps, prb, longbibliography]{revtex4-1}
\usepackage[dvips]{graphicx}
\usepackage{dcolumn}
\usepackage{bm}
\usepackage{color}

\begin{document}

\title{Nuclear spin-lattice relaxation in p-type GaAs}

\author{M.~Kotur}
\affiliation{Ioffe Institute, Russian Academy of Sciences, 194021 St-Petersburg, Russia}

\author{R. I. Dzhioev}
\affiliation{Ioffe Institute, Russian Academy of Sciences, 194021 St-Petersburg, Russia}

\author{M.~Vladimirova}
\affiliation{Laboratoire Charles Coulomb, UMR 5221 CNRS/ Universit\'{e}  de Montpellier,
F-34095, Montpellier, France}

\author{R.~V.~Cherbunin}
\affiliation{Spin Optics Laboratory, St-Petersburg State
University,  St-Peterbsurg, 198504, Russia}

\author{P.~S.~Sokolov}
\affiliation{Experimentelle Physik 2, Technische Universit\"at Dortmund, D-44221 Dortmund, Germany}
\affiliation{Spin Optics Laboratory, St-Petersburg State
University, St-Peterbsurg, 198504, Russia}

\author{D.~R.~Yakovlev }
\affiliation{Experimentelle Physik 2, Technische Universit\"at Dortmund, D-44221 Dortmund, Germany}

\affiliation{Ioffe Institute, Russian Academy of Sciences, 194021 St-Petersburg, Russia}

\author{M.~Bayer }
\affiliation{Experimentelle Physik 2, Technische Universit\"at Dortmund, D-44221 Dortmund, Germany}

\affiliation{Ioffe Institute, Russian Academy of Sciences, 194021 St-Petersburg, Russia}

\author{D.~Suter }
\affiliation{Experimentelle Physik 3, Technische Universit\"at Dortmund, D-44221 Dortmund, Germany}

\author{K.~V.~Kavokin}
\affiliation{Ioffe Institute, Russian Academy of Sciences, 194021 St-Petersburg, Russia}
\affiliation{Spin Optics Laboratory, St-Petersburg State
University, St-Peterbsurg, 198504, Russia}

\begin{abstract}
Spin-lattice relaxation of the nuclear spin system in p-type GaAs is studied using a three-stage experimental protocol including optical pumping and measuring the difference of the nuclear spin polarization before and after a dark interval of variable length. This method allows us to measure the spin-lattice relaxation time $T_1$ of optically pumped nuclei "in the dark", that is, in the absence of illumination. The measured $T_1$  values  fall into the sub-second time range, being three orders of magnitude shorter than in earlier studied n-type GaAs. The drastic difference is further emphasized by magnetic-field and temperature dependences of $T_1$ in p-GaAs, showing no similarity to those in n-GaAs. This unexpected behavior is explained within a developed theoretical model involving quadrupole relaxation of nuclear spins, which is induced by electric fields within closely spaced donor-acceptor pairs.

\end{abstract}

 \maketitle

%
\section{Introduction}
Optical pumping of nuclear spins via their dynamic polarization by photoexcited spin-polarized electrons is a powerful method for obtaining considerable nuclear polarization in semiconductors even in weak magnetic fields of the order of a few Gauss \cite{OpticalOrientation, DyakonovBook}. Creation and manipulation of the resulted Overhauser fields, acting upon the spins of charge carriers, presents multiple possibilities for studying the dynamics of mesoscopic spin systems. It is considered as one of the possible ways towards realization of spin-based information processing.
Gallium arsenide, a direct-bandgap semiconductor with the $100$ percent abundance of magnetic isotopes and strong hyperfine coupling, has been used as a test bench of the electron-nuclear spin dynamics since 1970s. It was known to specialists in the field (though, to the best of our knowledge, never explicitly mentioned in publications), that nuclear spin-lattice relaxation time $T_1$ in p-GaAs remains short even at liquid-helium temperatures, while n-GaAs demonstrates long $T_1$ (hundreds of seconds or even more) in this temperature range.

The spin-lattice relaxation of nuclei in n-GaAs was investigated in our recent works \cite{Kotur2014, Kotur2016,Vladimirova2017}. It was found to be dominated by the diffusion-limited hyperfine relaxation and quadrupole warm-up in lightly doped dielectric crystals, and by hyperfine relaxation involving both itinerant (Korringa mechanism) and localized electrons in heavily doped samples with metallic conductivity. In what concerns p-GaAs, even the time scale of the nuclear spin-lattice relaxation has not been exactly known.

In this paper, we present measurements of nuclear spin-lattice relaxation time $T_1$ as a function of magnetic field and temperature in two insulating p-GaAs layers with different concentrations of acceptors. The measured nuclear spin-lattice relaxation times are of the order of $100$~ms. They are independent of magnetic fields in the range $0-100$~G, and demonstrate a slow increase with lowering the temperature in the range $10-30$~K, which suddenly becomes sharp below $10$~K. These findings are drastically different from what is known about the nuclear spin relaxation in n-GaAs. The fact that nuclear spin relaxation in the dark, i.e. in the absence of photoexcited conduction-band electrons, is three orders of magnitude shorter than in n-GaAs, where resident electrons are abundant, is counterintuitive, since hyperfine coupling in the valence band is ten times weaker than in the conduction band  \cite{Grncharova,Fischer2008,Eble2009,Testelin2009,Fallahi2010}.  We propose a theoretical model that qualitatively explains the whole set of the experimental data for p-GaAs, and allows us to quantitatively reproduce the measured temperature dependence of nuclear spin relaxation time $T_1$.

\section{Samples and experimental setup}
\begin{figure}
\center{\includegraphics[width=1\linewidth]{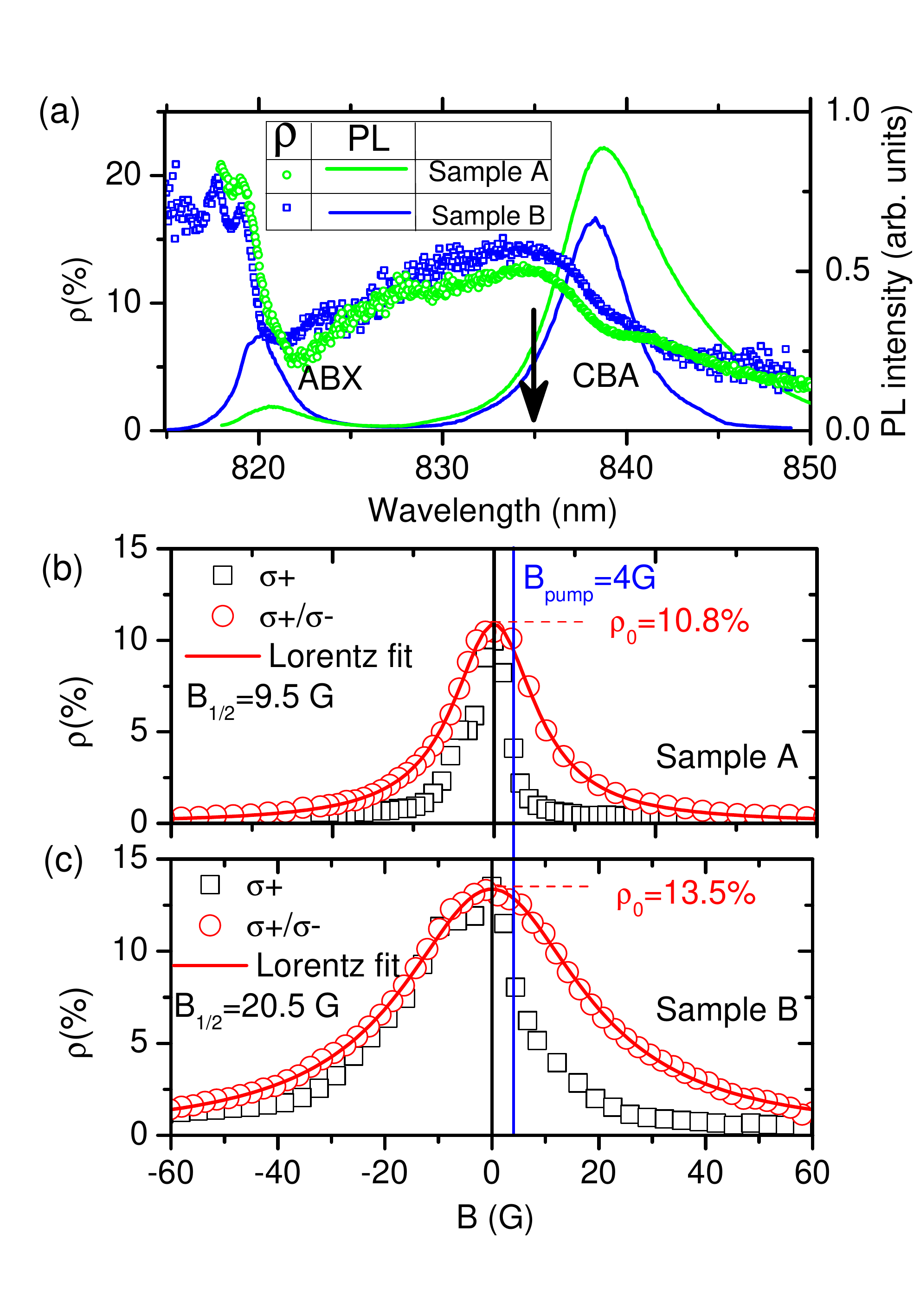} }
\caption{ (a) PL intensity  (right scale) and polarization (left scale) spectra for p-GaAs, Samples A and B
at $B=0$ and $T=5$~K. The two PL peaks are identified as acceptor-bound exciton (ABX) emission and conduction band-to-acceptor (CBA) recombination.
The black arrow indicates the chosen PL detection energy.
(b, c) PL polarization as a function of oblique magnetic field in Samples A (b) and B (c). Pump polarization is either alternated by a photoelastic modulator at the frequency of $50$~kHz (no Overhauser field, red symbols),
or fixed (black symbols). The onset of the Overhauser field results in the asymmetry with respect to zero. Solid lines are Lorentzian fits to the data, that allows for determination of 
$B_{1/2}$ and $\rho_0$ in Eq. (\ref{BN}).
}
\label{fig:spectra}
\end{figure}
\begin{figure*}
\center{\includegraphics[width=1\linewidth]{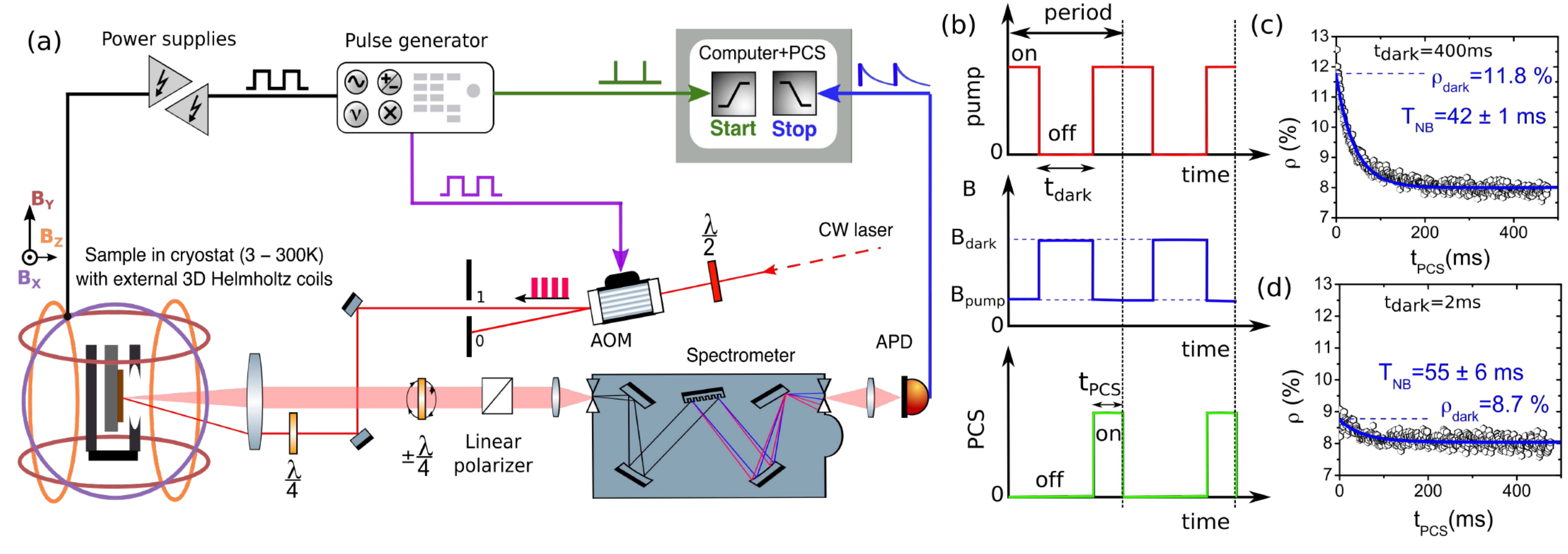} } \caption{
(a) Experimental setup designed for three stage measurements of sub-second nuclear spin relaxation times.
(b) Time-line of experiments. Two periods of pumping/magnetic field/photon counting sequence synchronously controlled by the pulse generator are shown ($100$ periods are used for each measurement at a given pump polarization).
(c, d) Typical PL polarization decays measured in Sample B (symbols) at $T=10$~K, $B=B_{dark}=4$~G, $t_{dark}=400$~ms (c) and  $t_{dark}=2$~ms (d). PL polarization $\rho_{dark}$ at the end of the dark interval ($t_{PCS}=0$) is recovered from the exponential fit (solid line) of the data. It is used to calculate the Overhauser field $B_N$ from Eq. (\ref{BN}) at given duration of dark interval and magnetic field. The decay time $T_{NB}$ of the exponential fit characterizes the nuclear spin relaxation time in the presence of optical pumping. It is always shorter than $T_1$.
}
\label{fig:setup}
\end{figure*}
The studied samples are two germanium-doped GaAs layers grown by liquid phase epitaxy on [001] GaAs substrate.
The corresponding acceptor concentrations are $n_A=2.6 \times10^{16}$ cm$^{-3}$ (Sample A) and $6.0 \times10^{16}$ cm$^{-3}$  (Sample B).
The samples are placed in a variable temperature cryostat (either helium flow or cold finger), surrounded by three pairs of Helmholtz coils. Such arrangement  allows for the compensation of the geomagnetic field and application of an external  mangnetic field in an arbitrary direction. Optical orientation of electron spins is achieved by pumping with a circularly ($\sigma^+$) polarized light from continuous wave (CW) titanium-sapphire laser at the wavelength $\lambda=800$~nm.  The light beam is directed along the sample axis. The spectra of photoluminescence (PL) intensity  and its circular polarization degree $\rho=(I^+-I^-)/(I^++I^-)$ for the two samples are shown in Fig.~\ref{fig:spectra}~(a). 
Here $I^+$ ($I^-$) is the intensity of PL emitted in $\sigma^+$ ($\sigma^-$) polarization, respectively.
 The two PL peaks can be identified as acceptor-bound exciton (ABX) emission and conduction band-to-acceptor (CBA) recombination.
Figs.~\ref{fig:spectra}~(b) and \ref{fig:spectra}~(c) present the PL polarization degree as a function of magnetic field $B$ applied at $80$ degrees with respect to the structure axis, for Samples A and B. When the pump polarization is alternated between $\sigma^+$ and $\sigma^-$ by a photoelastic modulator operating at the frequency of $50$~kHz (red curve), nuclear spins remain unpolarized, and $\rho(B)$ obeys the Lorentzian law (the Hanle effect) \cite{OpticalOrientation}. Under pumping by light with static circular polarization ($\sigma^+$), nuclear spins get polarized, and the Hanle curve is affected by the Overhauser field which is either parallel or anti-parallel to the external field, depending on the sign of the latter (black curve).

For studies of transient nuclear spin polarization $P_N$ we use the  experimental setup  shown in Fig.~\ref{fig:setup}~(a).
To asses millisecond time scale, we use pump pulses cut out of the CW laser beam with an acousto-optical modulator (AOM), controlled by the pulse generator. 
The same pulse generator controls the power supply of the 3D Helmholtz coils.
The PL polarization is measured in the reflection geometry at the spectral maximum of the PL polarization ($\lambda \approx835$~nm, see Fig. \ref{fig:spectra}~(a)). The emitted light passes through the spectrometer and is detected by the avalanche photodiode (APD), followed by the multi-channel photon counting system (PCS). The latter is synchronized with the AOM via the pulse generator.
The three stage experimental protocol that we implement here is very similar to that used in our previous work on n-GaAs \cite{Kotur2014,Kotur2016,Vladimirova2017},
but here it is adapted for measurement of subsecond nuclear spin relaxation times.
The  experiment timeline is shown in Fig. \ref{fig:setup}~(b).
The first stage of the experiment is the optical pumping of nuclei
 by circularly ($\sigma^+$) polarized light from a titanium-sapphire laser
 at $\lambda=800$~nm
  during $500$~ms.
The excitation power is $P_{pump}=4$~mW, focused on $90$~$\mu$m$^2$ spot on the sample surface.
The magnetic field $B_{pump}=4$~G is applied at $80$ degrees with respect to the structure axis.
At the second stage, the pump is switched off for an arbitrary time $t_{dark}$ (typically from $2$~ms to $1$~s), and the magnetic field is set to the value $B_{dark}$ at which we intend to measure the nuclear spin dynamics. $B_{dark}$ is parallel to $B_{pump}$ and ranges from zero to 120 G. The switching time is $\approx 1$~ms for $B<10$~G and $\approx 10$~ms for $B>10$~G).
At the end of the dark interval,  $B_{pump}$ is restored and the pump is switched on.
At the same moment the photon counting system starts the PL detection in either right or left circular polarization.
The PL signal is monitored during $500$~ms, which is sufficient to fully restore the nuclear spin polarization corresponding to the chosen pumping conditions.
At the end of this stage the cycle is repeated. The resulting PL signal is averaged over $100$ measurement cycles.
The same procedure is performed for the opposite polarization of PL.
From each pair of measurements the degree of circular polarization of PL is evaluated, and plotted  as a function of the photon counting time $t_{PCS}$.

\begin{figure}
\center{\includegraphics[width=1\linewidth]{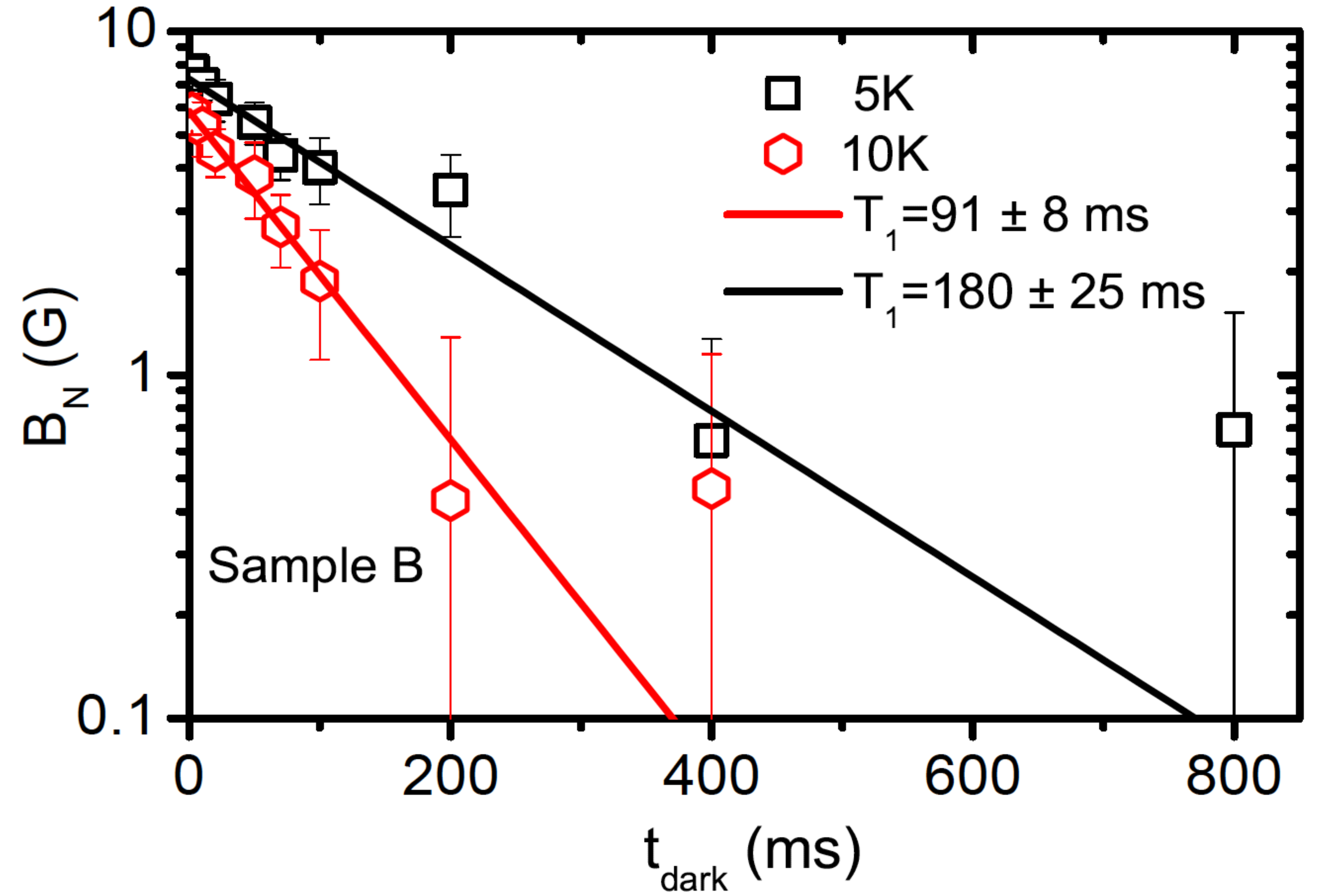} }
 \caption{Overhauser field $B_N$ derived using Eq.~(\ref{BN}) from PL polarization measurements as shown in Figs. \ref{fig:setup}(c, d) for various dark interval durations ($B=B_{dark}=4$~G). The exponential decay fit (solid lines) yields $T_1$ for  given temperatures.}
\label{fig:BN}
\end{figure}

Two examples of $\rho(t_{PCS})$ dependence for Sample B are presented in Figs. \ref{fig:setup}~(c) and \ref{fig:setup}~(d).  They correspond to dark intervals $t_{dark}=400$~ms and $2$~ms, respectively. One can see that $\rho$ decreases under  pumping down to $\rho_{pump}\approx 8\%$. This is a consequence of the chosen value of $B_{pump}=4$~G, at which the Overhauser field adds up to the external field and induces additional depolarization of electrons (see also  Fig. \ref{fig:spectra} (c), where the value of polarization under $\sigma^+$ pumping at $B=B_{pump}$ corresponds to $\rho=\rho_{pump}$). During dark intervals, nuclear polarization decreases, which results in a larger value of $\rho$ measured when the pump is back on. Fitting $\rho(t_{PCS})$ by exponential decay gives the build-up time of the nuclear field under optical pumping $T_{NB}$, as well as the value of $\rho_{dark}$. The knowledge of $\rho_{dark}$ allows for determination of the nuclear field intensity\cite{Kotur2014,Kotur2016}.
%
%
Indeed, the nuclear field can be recovered from the PL polarization degree using the following formula, derived from the well-known expression for the Hanle effect:
\begin{equation}
B_N= B_{1/2} \sqrt { \frac {\rho_0-\rho_{dark} }{\rho_{dark}}}-B_{pump}.
\label{BN}
\end{equation}
Here $\rho_{dark} $ is the degree of the PL polarization at the end of the dark interval [$\rho_{dark}\equiv \rho(t_{PCS}=0)$], $\rho_0$ is the PL polarization in the absence of the external field, and $B_{1/2}$ is the half width of the Hanle curve, measured independently  under conditions where nuclear spin polarization is absent (pump polarization modulated at $50$~kHz, see Figs.\ref{fig:spectra}~(b) and  \ref{fig:spectra}~(c)).
Even after shortest dark intervals, $B_N$ is a bit lower than before switching off the pump, most likely because of nuclear spin warm-up by the Knight field of photoexcited electrons, that rapidly changes when the pump is switched off and on \cite{Sokolov2017}.
By repeating the protocol for different durations of $t_{dark}$, we obtain $B_N$ relaxation curves for given values of temperature and applied magnetic field $B_{dark}$.
Examples of such dependences for two different temperatures are shown in Fig. \ref{fig:BN}.
 One can see that  $B_N$ decreases with increasing the length of the dark interval $t_{dark.}$. This  is due to nuclear spin-lattice relaxation "in the dark", that is in the absence of perturbation by pumping.
Exponential fitting of these curves yields the nuclear
spin relaxation time in the dark $T_1$, that we aim to study as a function of temperature and applied magnetic field $B_{dark}$.

%
\section{Experimental results and discussion}
\begin{figure}
\center{\includegraphics[width=1\linewidth]{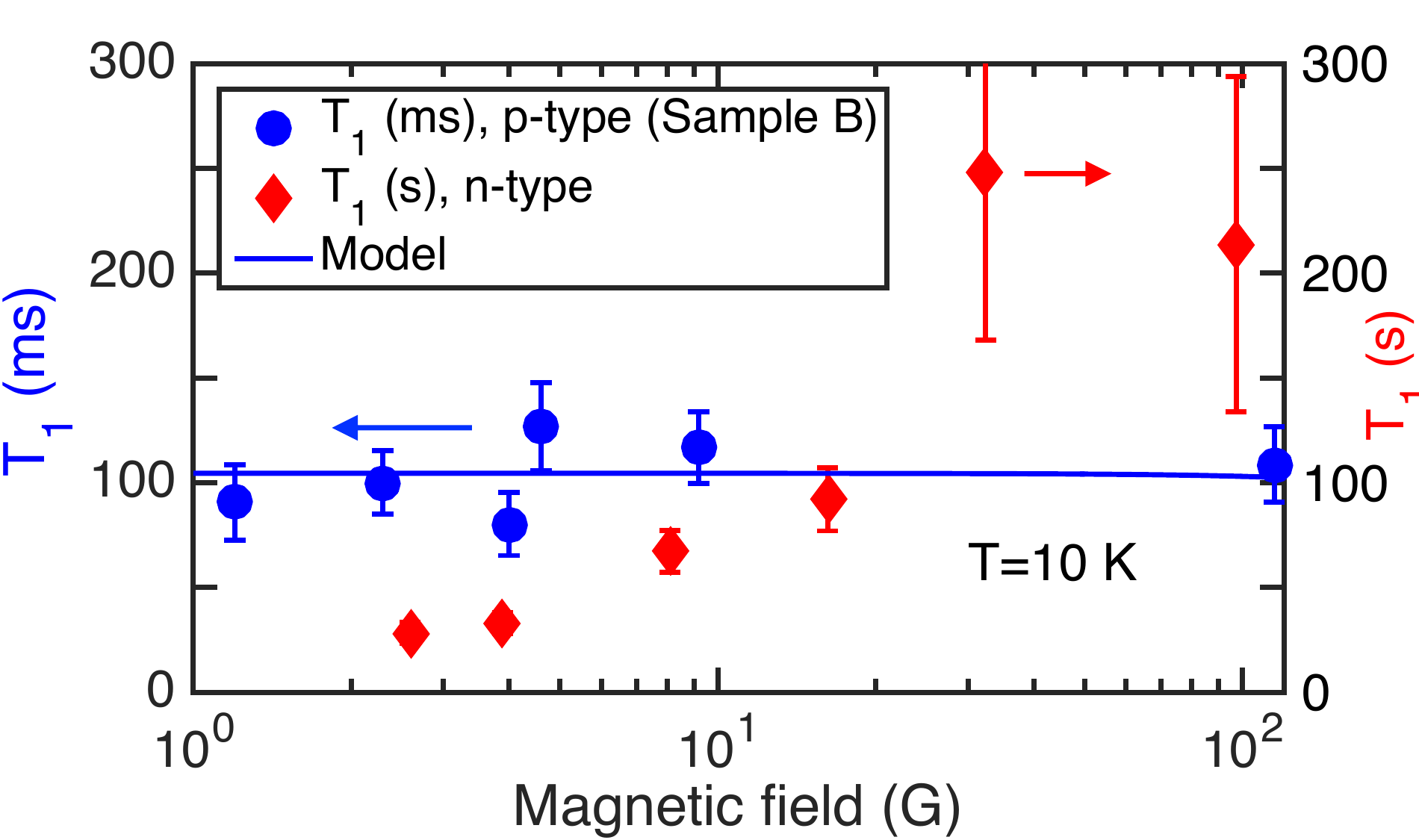} }
\caption{Magnetic field dependence of nuclear spin relaxation time.  Sample B (circles, left scale) and  n-GaAs with $n_D=6\times 10^{15}~\textnormal{cm}^{-3}$  (diamonds, right scale, data from Ref. \onlinecite{Vladimirova2017}). Solid line is a model prediction for p-GaAs, Eq.~(\ref{eq:T1B}).
}
\label{fig:Bdep}
\end{figure}
\begin{figure}
\center{\includegraphics[width=1\linewidth]{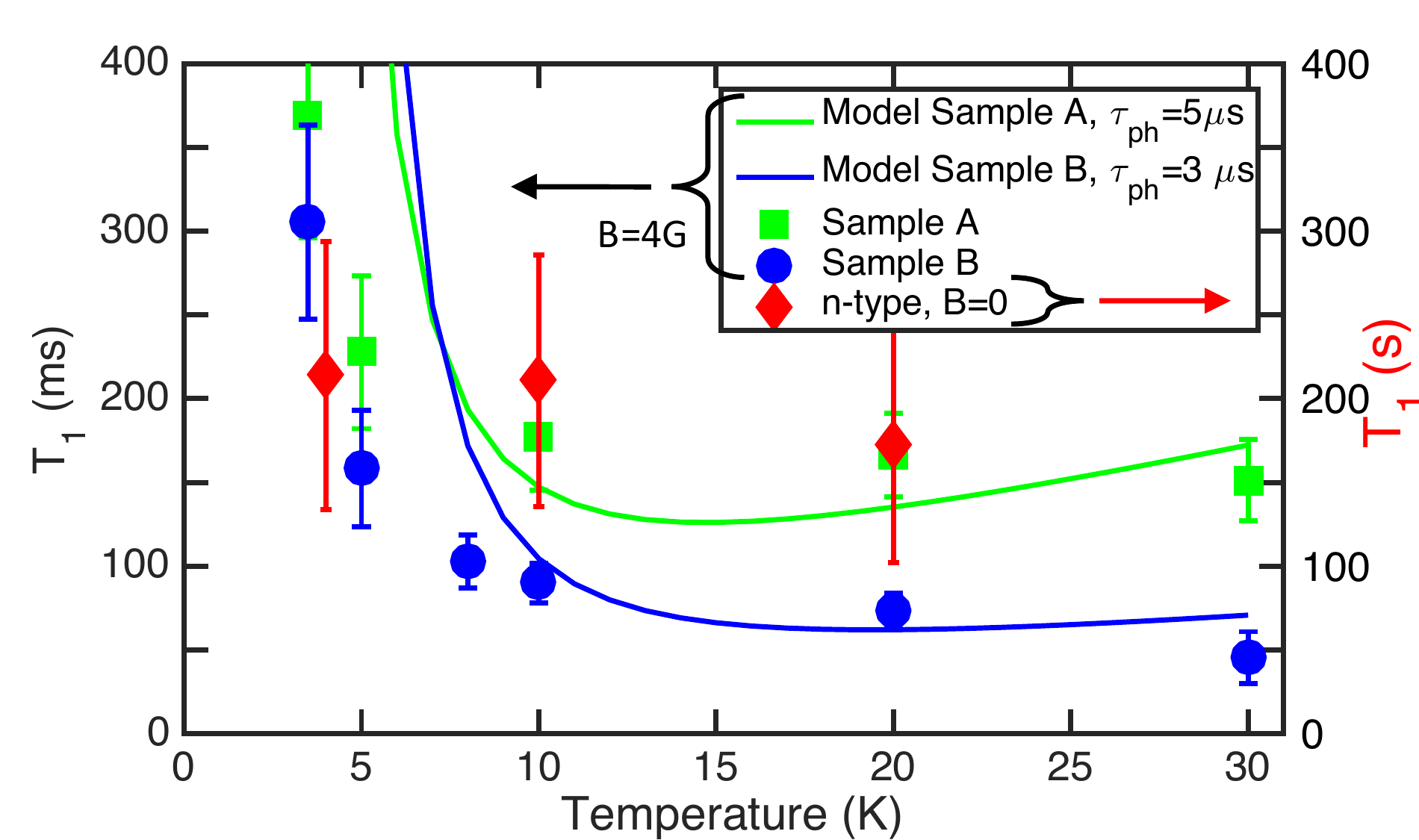} }
 \caption{Temperature dependence of nuclear spin relaxation time. In p-GaAs (left scale): Sample B (circles), Sample A (squares); in lightly doped n-GaAs  $n_D=6\times 10^{15}~\textnormal{cm}^{-3}$ (right scale:  diamonds, data from Ref. \onlinecite{Vladimirova2017}). Solid lines are model  
 predictions for  p-GaAs samples, calculated using Eq.~(\ref{eq:T1gen}).}
\label{fig:Tdep}
\end{figure}
Magnetic field dependence of nuclear spin relaxation time $T_1$ measured in Sample B  at $T=10$~K is shown in Fig. \ref{fig:Bdep}.
For comparison, the relaxation time in dielectric n-GaAs with donor concentration $n_D=6\times 10^{15}~\textnormal{cm}^{-3}$ (from Refs. \onlinecite{Vladimirova2017,Kotur2016}) is shown on the right scale.
One can see that in p-GaAs nuclear spin relaxation is about three orders of magnitude faster than in the n-GaAs.
Moreover, in n-GaAs the relaxation time exhibits a pronounced field dependence, while in p-GaAs it does not depend on the field.

The temperature dependence of $T_1$ is also surprising.
Figure~\ref{fig:Tdep} shows spin relaxation time measured in Samples A and B, as well as the comparison with n-GaAs sample with $n_D=6\times 10^{15}~\textnormal{cm}^{-3}$ from Refs. \onlinecite{Vladimirova2017,Kotur2016}.
It appears that in p-GaAs nuclear spin relaxation slows down significantly below $T=10$~K:  for example in Sample B $T_1=45$~ms at $T=30$~K and $310$~ms at $4$~K.
This behaviour is not observed in n-GaAs.
We discuss below possible mechanisms of nuclear spin lattice relaxation  in p-GaAs that could account for these experimental results,
taking advantage of the knowledge we already accumulated for n-GaAs.

The main feature distinguishing the spin-lattice relaxation of nuclei in p-GaAs from that in n-GaAs is its time scale, which is three orders of magnitude shorter. This fact excludes the diffusion-limited hyperfine relaxation \cite{Paget82, Vladimirova2017} from possible relaxation mechanisms. Indeed, with the nuclear spin diffusion constant $D \approx 10^{-13}$cm$^2$/s  the diffusion length during the time $100$~ms is just $1$~nm, i.e. two lattice constants of GaAs\cite{Paget82}. This means that the nuclear spin polarization does not have time to reach any remote killer center (e.g. paramagnetic impurity or neutral acceptor site) by diffusion, and decays within the same area where it has been created. Thus, spin diffusion that controls nuclear spin relaxation in n-GaAs  \cite{Paget82, Vladimirova2017} can not account for fast nuclear spin relaxation in p-GaAs.

This consideration leaves us two possible scenarios for  nuclear spin relaxation in p-GaAs:   either (i) nuclear spins are polarized only in regions where some efficient relaxation mechanism is at work, or (ii) a new, still unknown relaxation mechanism is acting everywhere in the crystal.

Let us start from the examination of the second scenario. So far, the only long-range relaxation mechanism known for the dielectric GaAs at low temperatures, is the quadrupole relaxation due to fluctuating electric fields.
Such fields result from hopping of localized charge carriers in the impurity band \cite{Kotur2016}.
However, this mechanism is far too weak to explain the observed relaxation time scale.
Indeed, the calculations reported in Ref. \onlinecite{Kotur2016} show that  one cannot expect the relaxation times shorter than $20$~s induced by this mechanism. This conclusion is corroborated by the experiments on n-GaAs.
Additionally, it was shown  that the efficiency of quadrupole relaxation of bulk nuclei drops down in magnetic fields exceeding the nuclear spin local field of the order of a few Gauss \cite{Kotur2016}. 
By contrast, in the studied p-GaAs samples no magnetic-field dependence of $T_1$ is observed at least up to $120$~G (Fig. \ref{fig:Bdep}).

\begin{figure}
 \begin{center}{}{ \includegraphics  [width=1\columnwidth]  {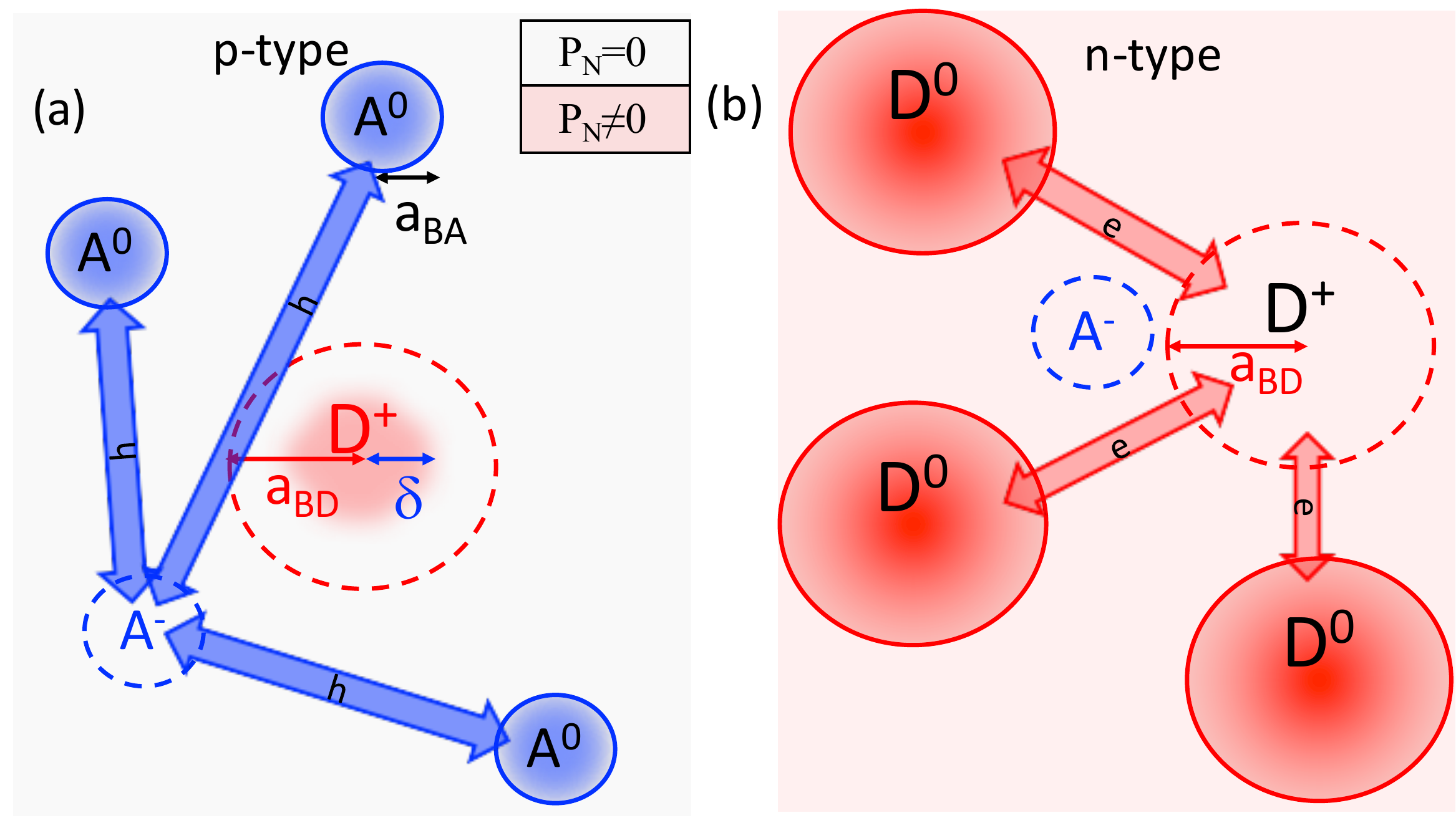} } 
 \end{center}
 \caption{Sketch of nuclear polarization patterns that form as a result of optical pumping in p-GaAs (a) and n-GaAs (b).  The degree of nuclear spin polarization $P_N$ is schematically represented by the red color intensity. 
In n-GaAs most of the donors are neutral; $P_N$ created under orbits of donor-bound electrons spreads into the inter-donor space so than $P_N\ne0$ everywhere under the light spot.  The relaxation of this polarization in the dark is provided by (i) diffusion towards neutral donors $D^0$, and (ii) directly in the inter-donor space via interaction of the nuclear quadrupole moments with the fluctuating electric field of hopping electrons  (between  $D^0$ and $D^+$). 
 In  p-GaAs all donors are charged; $P_N\ne0$ only near donors because of its quadrupole relaxation during pumping. Almost each donor has an acceptor nearby, and the electric charge of this acceptor fluctuates while it captures and releases a hole. This induces  quadrupole relaxation which is much faster in p-GaAs than in n-GaAs.
 }
 \label{fig:pn}
 \end{figure}

Exploring the first scenario, we note that the efficiency of dynamic polarization of nuclear spins by free photoexcited electrons is very low \cite{DYAKONOV1973}. Nuclear spins are polarized by electrons trapped to donor centers, which in p-type crystals are empty in the absence of optical excitation. Under continuous optical pumping,  the nuclear polarization can, in principle, spread out from the vicinity of donors into the bulk of the crystal by spin diffusion. This, indeed, occurs in n-type crystals. 
This is illustrated in Fig. \ref{fig:pn}~(b),  where a sketch of spatial distribution of the nuclear spin polarization in n-GaAs is shown. After sufficiently long pumping time nuclear spins are not only polarised within the donor orbits, but also
everywhere in the bulk, due to spin diffusion.  The relaxation of this polarization in the dark is then ensured by diffusion towards neutral donors $D^0$ (which play  here the role of the killer centers), and, at sufficiently low magnetic fields, directly in the bulk via interaction of the nuclear quadrupole moments with the fluctuating electric field of hopping charges.

\begin{figure}
\center{\includegraphics[width=01\linewidth]{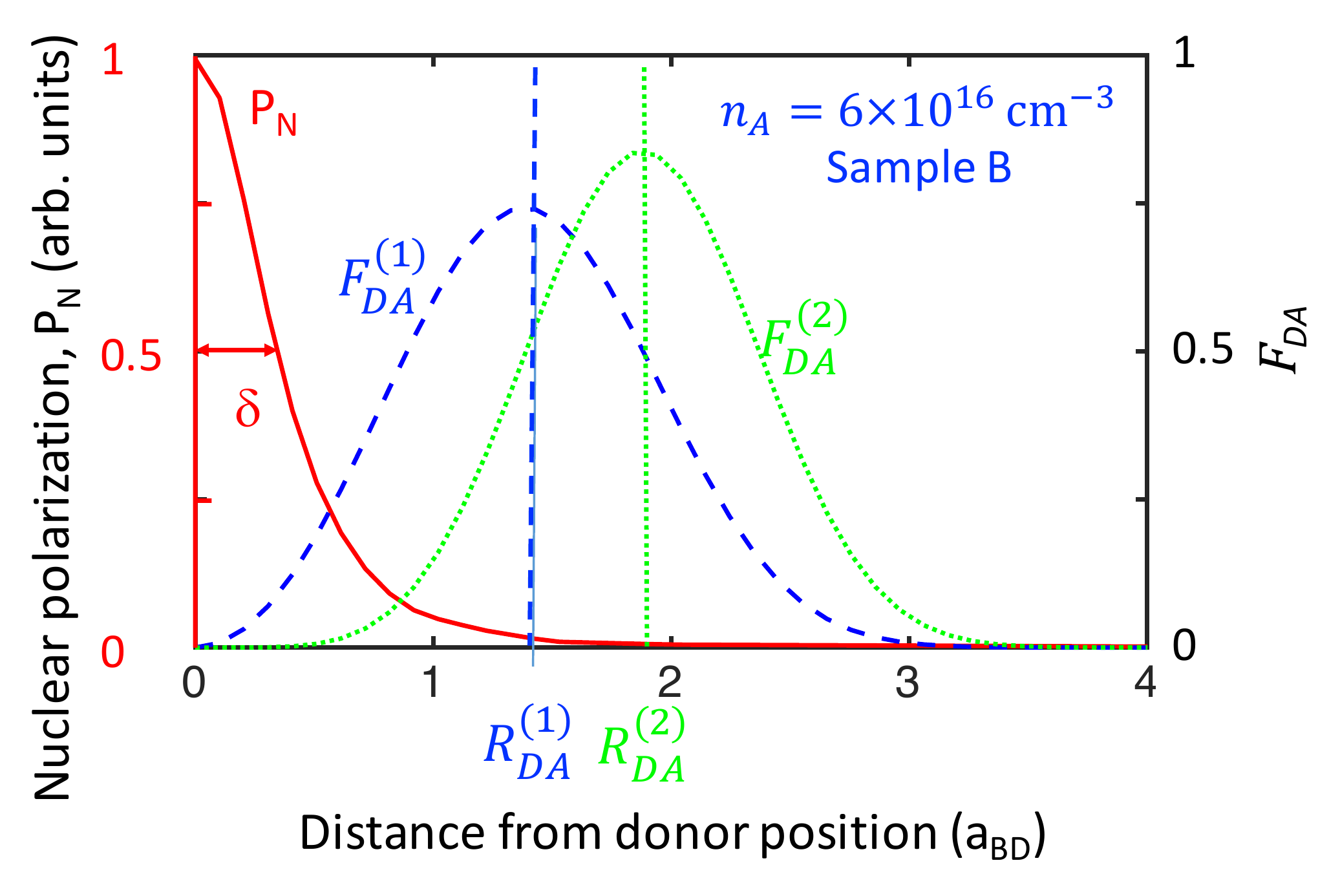} }
 \caption{Left scale: nuclear polarization created by optical pumping as a function of the distance from the donor (red solid line, from Ref. \onlinecite{PagetAmand}). Right scale: Probability density for the first (blue dashed line) and second (green dotted line) neighbouring acceptors calculated from Eq. (\ref{eq:F}) for Sample B as a function of the distance from the donor. The distance is expressed in the units of donor Bohr radius $a_{BD}=10$~nm.}
\label{fig:dist}
\end{figure}

In p-GaAs the situation is quite different. Let us first discuss how nuclear spin polaization in p-GaAs is generated under optical pumping. As shown by Paget, Amand and Korb \cite{PagetAmand}, in p-doped III-V semiconductors under optical pumping the nuclear polarization accumulates around donors only within the so-called "quadrupole radius" $\delta$, see Fig. \ref{fig:pn}~(a). It is defined by competition between hyperfine polarization and quadrupole relaxation of nuclei. The reason for this is as follows. Dynamic polarization of nuclei occurs due to their hyperfine coupling with the spins of electrons captured by donors, at the rate proportional to the electron spin density. The latter falls down exponentially with increasing the distance from the donor center \cite{Paget82}. Since photoexcited electrons spend at the donor only a limited time before recombination, the donor repeatedly changes its charge state from positively charged to neutral. This blinking charge creates a time-dependent electric field which, obeying the Coulomb law, extends far beyond the Bohr radius of the donor-bound electron $a_{BD}$. In piezoelectric semiconductors like GaAs, the electric field induces quadrupole splitting of nuclear spin states \cite{Brun}; fluctuating electric fields thus act similarly to fluctuating magnetic fields, causing nuclear spin relaxation. At some distance from the donor center, quadrupole relaxation overcomes dynamic polarization, because the electric field decreases with growing distance slower than the electron density.  According to the calculations reported in Ref. \onlinecite{PagetAmand}, this happens at approximately $0.4a_{BD}$. The numerically calculated dependence of the nuclear polarization on the distance from the donor is shown by red solid line in Fig. \ref{fig:dist}. Note, that the presence of an additional relaxation channel under pumping is corroborated by our experimental data: in both p-GaAs samples that we studied, the build-up time of the nuclear polarization is even shorter than its decay time in the dark, about $50$~ms, see Figs. \ref{fig:setup}~(c) and \ref{fig:setup}~(d).

\begin{figure}
\center{\includegraphics[width=1\linewidth]{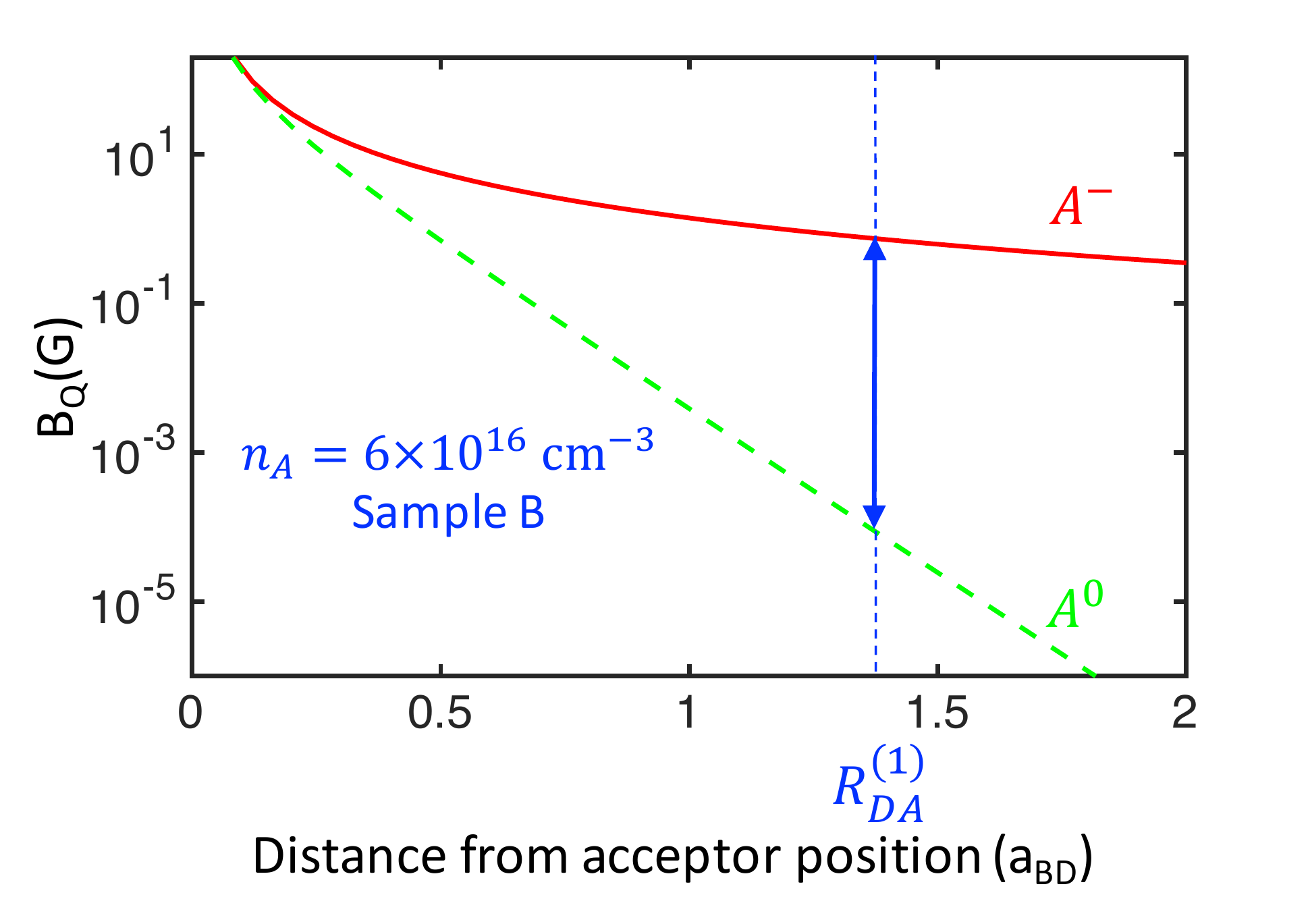} }
\caption{Effective quadrupole field $B_Q$ in the vicinity of the charged (red solid line) and neutral (green dashed line) acceptor. It is plotted as a function of the distance from the acceptor position. The distance is expressed in the units of donor Bohr radius $a_{BD}=10$~nm. Blue arrow shows that at the distance $R_{DA}^{(1)}$ corresponding to the maximum
probability to find the nearest donor  ({\it cf} Fig. \ref{fig:dist}), the charged acceptor  $A^-$ creates $B_Q \approx 1$~G, while $B_Q$ in the vicinity of $A^0$ is negligibly small.}
\label{fig:bq}
\end{figure}

Thus, we conclude that the nuclear polarization induced by optical pumping in p-GaAs is confined near donors, as sckethed in Fig. \ref{fig:pn}~(a). What relaxation mechanism can be responsible for its rapid decay in the dark, when the donor is empty? We suggest that it is the quadrupole relaxation induced by the electric field of a charged acceptor located in the vicinity of the donor. The presence of charged acceptors is a result of recombination of one of acceptor-bound holes with the donor electron. At zero temperature, the negative charge corresponding to the absence of a hole is located at the acceptor nearest to the positively charged donor with $97.4\%$ probability \cite{Shklovskii&EfrosCh3}. The distribution function of the distances from the donor to the nearest acceptor is shown in Fig.~\ref{fig:dist} by the blue dashed line (details of the corresponding calculation are given in Section \ref{model}). It has the maximum at $R^{(1)}_{DA}=(2\pi n_A)^{-1/3}$. For the studied range of $n_A$, this amounts to approximately $1.5 a_{BD}$. At this distance, a charged acceptor produces electric field $E$ of several kV/cm in the vicinity of the donor. Since GaAs is a polar crystal, this electric field induces an effective quadrupole field:
\begin{equation}
B_Q=b_QE,
 \end{equation}
where
\begin{equation}
b_Q=\frac{eQ\beta_Q}{4\gamma_N I(2I-1)}.
\label{eq:bQ}
 \end{equation}
$\beta_{Q}$ is the experimentally determined and isotope-dependent constant, $eQ$ is  the nuclear quadrupole moment, also isotope-dependent, $e$ is the absolute value of the electron charge, $\gamma_N$ is the nuclear gyromagnetic ratio, $I$ is the nuclear spin. \cite{Brun,PagetAmand,Petrov}.
For GaAs  $b_Q\approx 0.2$~G$\cdot$cm/kV. Figure~\ref{fig:bq} shows the  effective quadrupole field as a function of the distance from the charged (red solid line) and neutral (green dashed line) acceptor.
One can see that at the distance $R^{(1)}_{DA}$ from the $A^-$ acceptor $B_Q \approx 1$~G.
%

If $T \ne 0$, a hole from a more remote, neutral, acceptor can jump to this site, neutralizing it. Thus, fluctuations of the occupation number of the nearest acceptor produce fluctuating quadrupole fields.  In Section \ref{model} we show that  these fluctuations provide an efficient nuclear spin relaxation.

Figure~\ref{fig:pn} summarises the above considerations via comparison of nuclear polarization patterns that form as a result of optical pumping in p-GaAs and n-GaAs. In n-GaAs, most of the donors are neutral; nuclear polarization created under orbits of donor-bound electrons spreads into the inter-donor space. The number of charged donor-acceptor pairs is small and most of nuclei are situated far from such pairs.
 To the opposite, in p-GaAs all the donors are charged; nuclear polarization is concentrated near donors because of its quadrupole relaxation during pumping \cite{PagetAmand}. Almost each donor has an acceptor nearby, and the electric charge of this acceptor fluctuates while it captures and releases a hole. This explains why quadrupole nuclear-spin relaxation induced by fluctuating charges is three orders of magnitude faster in p-GaAs as compared to n-GaAs.

This model explains also the $T_1$ behavior as a function of temperature and magnetic field in p-GaAs.
Indeed, with lowering temperature the fraction of time, during which the nearest acceptor site is charged, increases, since this state is energetically favourable. As a result, charge distribution in the vicinity of the donor becomes frozen, and the electric field stops fluctuating. This obviously should lead to an increase of $T_1$, and this is exactly what is observed in experiments, see Fig. \ref{fig:Tdep}.
%

The $T_1$ independence of the applied magnetic field $B$ means that $\omega_B\tau_c<<1$, where $\omega_B=\gamma_NB$ is the nuclear Larmor frequency in the field $B$  and $\tau_c$ is the correlation time of the fluctuating field which causes the spin relaxation  \cite{DyakonovBook}.
For nuclear species of GaAs, the average nuclear gyromagnetic ratio $\langle \gamma_N\rangle \approx 9\times 10^3$~rad/G$\cdot$s.
Using the well-known formula for spin relaxation under influence of a fluctuating magnetic field \cite{DyakonovBook}:
\begin{equation}
 \frac{1}{T_1}={\omega_f}^2 \tau_c,
 \label{eq:t1}
 \end{equation}
where $\omega_f=\gamma_NB_f$ is the spin precession frequency in the fluctuating field $B_f$, one can estimate what correlation time would explain the observed $T_1 \approx 100$~ms (see Fig.~\ref{fig:Bdep}).
Assuming the magnitude of the fluctuating quadrupole field of $1$~G as estimated above (see Fig. \ref{fig:bq} and details of the calculations in Section \ref{model}) we get
 $\tau_c \approx 100$~ns.
Therefore, throughout the range of magnetic fields applied in our experiment, the condition $\omega_B\tau_c<<1$ is satisfied,
which is consistent with $T_1$ independence of the magnetic field, up to the maximum  magnetic field $B=120$~G applied in our experiments.
In Section \ref{model} we present the theoretical model which quantifies the above considerations.


\section{Theoretical model}
\label{model}
Let us assume that the fluctuations of the electric field experienced by the optically polarized nuclei under the donor orbit
result from the charge fluctuations on the nearest acceptor, as shown in Fig. \ref{fig:pn}.
When it is negatively charged, it creates an electric field ${\boldsymbol{E}}_{-}$ in the vicinity of the donor (we neglect spatial variation of this field within the sphere with the radius $\delta$, where nuclear spins are polarized). When the nearest acceptor is neutral, we assume that the electric field takes certain value $\it{E}_{0}$; in doing so, we neglect the variability of charge configuration at more remote impurities.

We denote the average time during which the nearest acceptor stays charged as $\tau_{-}$, and the average time during which it is neutral as $\tau_{0}$. The average electric field at the donor is then equal to

\begin{equation}
 \langle \boldsymbol{E} \rangle=\frac{\boldsymbol{E}_{-}\tau_{-}+\boldsymbol{E}_{0}\tau_{0}}{\tau_{-}+\tau_{0}}.
 \end{equation}
The mean squared fluctuation of this field is, correspondingly,
\begin{equation}
 \delta E^2=\langle (\boldsymbol{E}-\langle \boldsymbol{E} \rangle)^2 \rangle=
 \Delta E^2 \frac{\tau_{-}\tau_{0}}{(\tau_{-}+\tau_{0})^2},
 \label{eq:mean}
 \end{equation}
where  $\Delta E^2 \equiv (\boldsymbol{E}_{-}-\boldsymbol{E}_{0})^2$.    The autocorrelation function of the fluctuating part of the electric field, $\boldsymbol{E}_{f}=\boldsymbol{E}-\langle \boldsymbol{E} \rangle$, which presents an example of an asymmetric random telegraph signal, is determined by the shortest of the two times, $\tau_{-}$ and $\tau_{0}$:
\begin{equation}
 \langle \boldsymbol{E}(t) \cdot \boldsymbol{E}(0) \rangle=
  \delta E^2 \exp{\Big(-t(\frac{\tau_{-}+\tau_{0}}{\tau_{-}\tau_{0}})\Big)}.
   \label{eq:autoc}
 \end{equation}
 In other words, the correlation time of the electric field fluctuations is equal to
 \begin{equation}
 \tau_c=\frac{\tau_{-}\tau_{0}}{\tau_{-}+\tau_{0}}.
 \end{equation}
 The Fourier transform of the autocorrelation function in Eq.~(\ref{eq:autoc}) allows calculating the spectral power density of electric field fluctuations at the donor:
 \begin{equation}
 \delta E^2_{\omega}=
 \frac{\delta E^2 \tau_c }{1+\omega^2(\tau_{-}\tau_{0})^2/(\tau_{-}+\tau_{0})^2}.
 \end{equation}
Therefore,  the resulting spectral power density of the quadrupole-induced effective magnetic field is given by
 \begin{equation}
 \delta B^2_{\omega}= b^2_Q \delta B^2_{\omega}.
 \end{equation}
 %

According to  Abragam \cite{Abragam}  the spin relaxation rate of the nuclear spin system in presence of the fluctuating magnetic field reads:
\begin{equation}
\frac{1}{T_1}=\gamma_N^2 \delta B_{\omega}.
\label{eq:abragam}
\end{equation}
At Larmor frequency of nuclear spin in the external field $B$, $\omega=\omega_B$.
At low magnetic fields used in this study and satisfying the condition $\omega_B \tau_c <<1$,
Eq.~(\ref{eq:abragam}) reduces to Eq.~(\ref{eq:t1}).
%
%
%
%
%
 %
Thus we arrive to the following expression for $T_1^{-1}$:

\begin{equation}
\frac{1}{T_1}\approx
 \gamma_N^2 b_Q^2 \Delta E^2 \frac{(\tau_{-}\tau_{0})^2}{(\tau_{-}+\tau_{0})^3}.
 \end{equation}
The characteristic times $\tau_{-}$ and $\tau_{0}$ are determined by probabilities of phonon-assisted transitions between the configurations with charged and neutral nearest acceptor.
The most reasonable assumption, that can be done to estimate these times, is to suppose that these transitions correspond to the hopping of a hole between two acceptors closest to the donor, see Fig. \ref{fig:pn}~(a).
%
%
Denoting the hole energy at the nearest acceptor by $\epsilon_{-}$,  and at the second nearest acceptor as $\epsilon_{0}$ we arrive to the following expressions for $\tau_{-}$ and $\tau_{0}$:

\begin{eqnarray}
 \tau_{-}=\frac{\tau_{ph}}{n_{ph}} ;
 \tau_{0}=\frac{\tau_{ph}}{n_{ph}+1},
 \end{eqnarray}
 where $\tau_{ph}$ is the characteristic time of the corresponding phonon-assisted transition, $n_{ph}$ is the number of phonons given by the Planck distribution:
\begin{equation}
 n_{ph}=\frac{1}{\exp\left[\Delta \epsilon/ (k_B T)\right]+1},
 \end{equation}
 and $\Delta \epsilon =\epsilon_{-}-\epsilon_0$.
 Finally we obtain:
\begin{eqnarray}
\frac{1}{T_1}\approx
 \gamma_N^2 b_Q^2 \Delta E^2  \tau_{ph}\times \nonumber \\
 \frac{\left[1-\exp{(-\Delta \epsilon /k_B T)}\right ]\exp{(-\Delta \epsilon /k_B T)}}{\left[1+\exp{(-\Delta \epsilon /k_B T)} \right]^3}.
 \label{eq:T1gen}
 \end{eqnarray}

Eq.~(\ref{eq:T1gen}) allows one to calculate the temperature dependence of $T_1$.
In order to do that, we need to estimate $\Delta \epsilon$ and $\Delta E^2$.
They are given by the Coulomb energies and electric fields of two charges located at distances $r_{DA}^{(1)}$ and $r_{DA}^{(2)}$
from the donor to two nearest acceptors.
As an estimate for these distances we take the maxima of the first and second neighbouring acceptor distributions:
\begin{eqnarray}
 F_{DA}^{(1)}=4 \pi r_{DA}^2
n_A \exp{(-\frac{4}{3}\pi r_{DA}^3n_A)} ;\nonumber \\
 F_{DA}^{(2)}=\frac{16}{3} \pi r_{DA}^2
n_A \exp{(-\frac{4}{3}\pi^2 r_{DA}^5n_A^2)}.
\label{eq:F}
  \end{eqnarray}
These distributions are shown
in Fig.~\ref{fig:dist} for Sample B.
The maxima of these distributions are given by
\begin{eqnarray}
 R_{DA}^{(1)}=(2 \pi n_A)^{-1/3};  \nonumber \\
R_{DA}^{(2)}=(\frac{4 \pi}{5} n_A)^{-1/3}.
  \end{eqnarray}
Thus, we obtain
\begin{eqnarray}
\Delta \epsilon= -\frac{e^2}{4 \pi \epsilon \epsilon_0} 
\left[ \frac{1}{R_{DA}^{(1)}}-\frac{1}{R_{DA}^{(2)}} \right ] ; \nonumber \\
\Delta E^2=\left( \frac{e}{4 \pi \epsilon \epsilon_0} \right) ^2\left [  \left ( \frac{1}{R_{DA}^{(1)}}\right )^4+  \left ( \frac{1}{R_{DA}^{(2)}}\right )^4 \right ]
\label{eq:eps}
  \end{eqnarray}
where we averaged the squared electric field over angular distribution of the two acceptors.

Eq. (\ref{eq:T1gen}) together with Eqs. (\ref{eq:bQ}) and (\ref{eq:eps}) leaves us with the only fitting parameter, $\tau_{ph}$, to reproduce the measured low-field temperature dependence of the nuclear spin relaxation time shown in Fig. \ref{fig:Tdep}.
The suppression of the spin relaxation by application of the magnetic field can be  calculated from this value of $T_1$  using the motional narrowing formula \cite{DyakonovBook}:
\begin{equation}
 T_1(B)=\frac{T_1}{1+\omega_B^2 \tau_c^2}.
 \label{eq:T1B}
  \end{equation}
The results of the fitting procedure are shown in Figs. \ref{fig:Bdep} and  \ref{fig:Tdep} by solid lines. One can see that the agreement is quite reasonable: there is no suppression of the nuclear spin relaxation up to $120$~G (no magnetic field dependence) and the quenching of spin relaxation below $T\approx 10$~K is well reproduced assuming $\tau_{ph}=5$~$\mu$s in Sample A and $\tau_{ph}=3$~$\mu$s in Sample B.

We note that $\tau_{ph}$ obtained by fitting experimental data  yields a qualitatively correct trend as a function of acceptor concentration (phonon-assisted hops become more frequent with decreasing the average distance between nearest acceptors). However, since the overlap of wave functions of impurity-bound holes decreases exponentially with growing distance, one would expect a greater difference in $\tau_{ph}$ between the two studied p-GaAs samples. This might be an indication that our model, which takes into account only two nearest acceptors, is too simplistic. In order to clarify this issue, an extensive experimental study of nuclear spin relaxation over a broad range of doping in p-GaAs is needed. Such studies could be an interesting subject for the future work.


\section{Conclusions}
In conclusion, we have investigated  relaxation of nuclear spin polarization created by optical pumping in bulk p-GaAs. The nuclear spin-lattice relaxation time $T_1$ in the dark (that is in the absence of optical pumping) turns out to be longer than that under pumping, but still remains in the sub-second range.  This is three orders of magnitude shorter than in n-GaAs. This fact seems counterintuitive, since hyperfine coupling of holes is much weaker than that of conduction-band electrons. This paradox can be solved by taking into account charge fluctuations at acceptors located in close vicinity of positively charged donor centers. Indeed, since optically induced nuclear spin polarization is created only in the vicinity of donors and cannot diffuse outwards, the nearby fluctuating charge effectively destroys the nuclear polarization via the quadrupole interaction. The proposed theoretical model quantitatively describes the slowing down of nuclear spin relaxation below $T=10$~K (due to slowing down of charge fluctuations), and magnetic-field independence (up to $\approx 100$~G) of $T_1$.
Our results fill the gap in the general picture of nuclear spin relaxation in doped GaAs, where p-type doping
has not been addressed so far. They also suggest that in compounds with $I=1/2$, like p-CdTe, nuclear spin relaxation can be much slower, due to the absence of the quadrupole effects.
%

\emph{Acknowledgements}.  This work was supported
by the Russian Foundation for Basic Research (RFBR, Grants No. 16-52-150008 and 15-52-12020),
by the Ministry of Education and Science of the Russian Federation (contract
No. 14.Z50.31.0021 with the Ioffe Institute, Russian Academy of Sciences,
and leading researcher M. Bayer),
by Saint-Petersburg State University via a research grant 11.34.2.2012,
by French National Research Agency (Grant  OBELIX, No. ANR-15-CE30-0020-02) and
National Center for Scientific Research (CNRS, PRC SPINCOOL No. 148362),
and by Deutsche Forschungsgemeinschaft in the frame of the ICRC TRR
160 (Project No. A6).


\begin{thebibliography}{18}%
\makeatletter
\providecommand \@ifxundefined [1]{%
 \@ifx{#1\undefined}
}%
\providecommand \@ifnum [1]{%
 \ifnum #1\expandafter \@firstoftwo
 \else \expandafter \@secondoftwo
 \fi
}%
\providecommand \@ifx [1]{%
 \ifx #1\expandafter \@firstoftwo
 \else \expandafter \@secondoftwo
 \fi
}%
\providecommand \natexlab [1]{#1}%
\providecommand \enquote  [1]{``#1''}%
\providecommand \bibnamefont  [1]{#1}%
\providecommand \bibfnamefont [1]{#1}%
\providecommand \citenamefont [1]{#1}%
\providecommand \href@noop [0]{\@secondoftwo}%
\providecommand \href [0]{\begingroup \@sanitize@url \@href}%
\providecommand \@href[1]{\@@startlink{#1}\@@href}%
\providecommand \@@href[1]{\endgroup#1\@@endlink}%
\providecommand \@sanitize@url [0]{\catcode `\\12\catcode `\$12\catcode
  `\&12\catcode `\#12\catcode `\^12\catcode `\_12\catcode `\%12\relax}%
\providecommand \@@startlink[1]{}%
\providecommand \@@endlink[0]{}%
\providecommand \url  [0]{\begingroup\@sanitize@url \@url }%
\providecommand \@url [1]{\endgroup\@href {#1}{\urlprefix }}%
\providecommand \urlprefix  [0]{URL }%
\providecommand \Eprint [0]{\href }%
\providecommand \doibase [0]{http://dx.doi.org/}%
\providecommand \selectlanguage [0]{\@gobble}%
\providecommand \bibinfo  [0]{\@secondoftwo}%
\providecommand \bibfield  [0]{\@secondoftwo}%
\providecommand \translation [1]{[#1]}%
\providecommand \BibitemOpen [0]{}%
\providecommand \bibitemStop [0]{}%
\providecommand \bibitemNoStop [0]{.\EOS\space}%
\providecommand \EOS [0]{\spacefactor3000\relax}%
\providecommand \BibitemShut  [1]{\csname bibitem#1\endcsname}%
\let\auto@bib@innerbib\@empty
\bibitem [{\citenamefont {Meier}\ and\ \citenamefont
  {Zakharchenya}(1984)}]{OpticalOrientation}%
  \BibitemOpen
  \bibinfo {editor} {\bibfnamefont {F.}~\bibnamefont {Meier}}\ and\ \bibinfo
  {editor} {\bibfnamefont {B.}~\bibnamefont {Zakharchenya}},\ eds.,\ \href@noop
  {} {\emph {\bibinfo {title} {Optical Orientation}}},\ \bibinfo {series}
  {Modern Problems in Condensed Matter Science Series}, Vol.~\bibinfo {volume}
  {8}\ (\bibinfo  {publisher} {North-Holland, Amsterdam},\ \bibinfo {year}
  {1984})\BibitemShut {NoStop}%
\bibitem [{\citenamefont {{Dyakonov}}(2017)}]{DyakonovBook}%
  \BibitemOpen
  \bibinfo {editor} {\bibfnamefont {M.~I.}\ \bibnamefont {{Dyakonov}}},\ ed.,\
  \href {\doibase 10.1007/978-3-319-65436-2} {\emph {\bibinfo {title} {Spin
  Physics in Semiconductors}}},\ Springer series in solid-state sciences\
  (\bibinfo  {publisher} {Springer International Publishing},\ \bibinfo {year}
  {2017})\BibitemShut {NoStop}%
\bibitem [{\citenamefont {Kotur}\ \emph {et~al.}(2014)\citenamefont {Kotur},
  \citenamefont {Dzhioev}, \citenamefont {Kavokin}, \citenamefont {Korenev},
  \citenamefont {Namozov}, \citenamefont {Pak},\ and\ \citenamefont
  {Kusrayev}}]{Kotur2014}%
  \BibitemOpen
  \bibfield  {author} {\bibinfo {author} {\bibfnamefont {M.}~\bibnamefont
  {Kotur}}, \bibinfo {author} {\bibfnamefont {R.~I.}\ \bibnamefont {Dzhioev}},
  \bibinfo {author} {\bibfnamefont {K.~V.}\ \bibnamefont {Kavokin}}, \bibinfo
  {author} {\bibfnamefont {V.~L.}\ \bibnamefont {Korenev}}, \bibinfo {author}
  {\bibfnamefont {B.~R.}\ \bibnamefont {Namozov}}, \bibinfo {author}
  {\bibfnamefont {P.~E.}\ \bibnamefont {Pak}}, \ and\ \bibinfo {author}
  {\bibfnamefont {Yu.~G.}\ \bibnamefont {Kusrayev}},\ }\bibfield  {title}
  {\enquote {\bibinfo {title} {{Nuclear spin relaxation mediated by Fermi-edge
  electrons in n-type GaAs}},}\ }\href@noop {} {\bibfield  {journal} {\bibinfo
  {journal} {JETP Lett.}\ }\textbf {\bibinfo {volume} {99}},\ \bibinfo {pages}
  {37} (\bibinfo {year} {2014})}\BibitemShut {NoStop}%
\bibitem [{\citenamefont {Kotur}\ \emph {et~al.}(2016)\citenamefont {Kotur},
  \citenamefont {Dzhioev}, \citenamefont {Vladimirova}, \citenamefont
  {Jouault}, \citenamefont {Korenev},\ and\ \citenamefont
  {Kavokin}}]{Kotur2016}%
  \BibitemOpen
  \bibfield  {author} {\bibinfo {author} {\bibfnamefont {M.}~\bibnamefont
  {Kotur}}, \bibinfo {author} {\bibfnamefont {R.~I.}\ \bibnamefont {Dzhioev}},
  \bibinfo {author} {\bibfnamefont {M.}~\bibnamefont {Vladimirova}}, \bibinfo
  {author} {\bibfnamefont {B.}~\bibnamefont {Jouault}}, \bibinfo {author}
  {\bibfnamefont {V.~L.}\ \bibnamefont {Korenev}}, \ and\ \bibinfo {author}
  {\bibfnamefont {K.~V.}\ \bibnamefont {Kavokin}},\ }\bibfield  {title}
  {\enquote {\bibinfo {title} {{Nuclear spin warm up in bulk n-GaAs}},}\
  }\href@noop {} {\bibfield  {journal} {\bibinfo  {journal} {Phys. Rev. B}\
  }\textbf {\bibinfo {volume} {94}},\ \bibinfo {pages} {081201} (\bibinfo
  {year} {2016})}\BibitemShut {NoStop}%
\bibitem [{\citenamefont {Vladimirova}\ \emph {et~al.}(2017)\citenamefont
  {Vladimirova}, \citenamefont {Cronenberger}, \citenamefont {Scalbert},
  \citenamefont {Kotur}, \citenamefont {Dzhioev}, \citenamefont {Ryzhov},
  \citenamefont {Kozlov}, \citenamefont {Zapasskii}, \citenamefont
  {Lema{\^\i}tre},\ and\ \citenamefont {Kavokin}}]{Vladimirova2017}%
  \BibitemOpen
  \bibfield  {author} {\bibinfo {author} {\bibfnamefont {M.}~\bibnamefont
  {Vladimirova}}, \bibinfo {author} {\bibfnamefont {S.}~\bibnamefont
  {Cronenberger}}, \bibinfo {author} {\bibfnamefont {D.}~\bibnamefont
  {Scalbert}}, \bibinfo {author} {\bibfnamefont {M.}~\bibnamefont {Kotur}},
  \bibinfo {author} {\bibfnamefont {R.~I.}\ \bibnamefont {Dzhioev}}, \bibinfo
  {author} {\bibfnamefont {I.~I.}\ \bibnamefont {Ryzhov}}, \bibinfo {author}
  {\bibfnamefont {G.~G.}\ \bibnamefont {Kozlov}}, \bibinfo {author}
  {\bibfnamefont {V.~S.}\ \bibnamefont {Zapasskii}}, \bibinfo {author}
  {\bibfnamefont {A.}~\bibnamefont {Lema{\^\i}tre}}, \ and\ \bibinfo {author}
  {\bibfnamefont {K.~V.}\ \bibnamefont {Kavokin}},\ }\bibfield  {title}
  {\enquote {\bibinfo {title} {{Nuclear spin relaxation in n-GaAs: From
  insulating to metallic regime}},}\ }\href@noop {} {\bibfield  {journal}
  {\bibinfo  {journal} {Phys. Rev. B}\ }\textbf {\bibinfo {volume} {95}},\
  \bibinfo {pages} {125312} (\bibinfo {year} {2017})}\BibitemShut {NoStop}%
\bibitem [{\citenamefont {Grncharova}\ and\ \citenamefont
  {Perel}(1977)}]{Grncharova}%
  \BibitemOpen
  \bibfield  {author} {\bibinfo {author} {\bibfnamefont {E~I}\ \bibnamefont
  {Grncharova}}\ and\ \bibinfo {author} {\bibfnamefont {V~I}\ \bibnamefont
  {Perel}},\ }\bibfield  {title} {\enquote {\bibinfo {title} {{Spin relaxation
  in semiconductors caused by electric fields}},}\ }\href@noop {} {\bibfield
  {journal} {\bibinfo  {journal} {Sov. Phys. Semicond.}\ }\textbf {\bibinfo
  {volume} {11}},\ \bibinfo {pages} {997} (\bibinfo {year} {1977})}\BibitemShut
  {NoStop}%
\bibitem [{\citenamefont {Fischer}\ \emph {et~al.}(2008)\citenamefont
  {Fischer}, \citenamefont {Coish}, \citenamefont {Bulaev},\ and\ \citenamefont
  {Loss}}]{Fischer2008}%
  \BibitemOpen
  \bibfield  {author} {\bibinfo {author} {\bibfnamefont {Jan}\ \bibnamefont
  {Fischer}}, \bibinfo {author} {\bibfnamefont {W.~A.}\ \bibnamefont {Coish}},
  \bibinfo {author} {\bibfnamefont {D.~V.}\ \bibnamefont {Bulaev}}, \ and\
  \bibinfo {author} {\bibfnamefont {Daniel}\ \bibnamefont {Loss}},\ }\bibfield
  {title} {\enquote {\bibinfo {title} {{Spin decoherence of a heavy hole
  coupled to nuclear spins in a quantum dot}},}\ }\href@noop {} {\bibfield
  {journal} {\bibinfo  {journal} {Phys. Rev. B}\ }\textbf {\bibinfo {volume}
  {78}},\ \bibinfo {pages} {155329} (\bibinfo {year} {2008})}\BibitemShut
  {NoStop}%
\bibitem [{\citenamefont {Eble}\ \emph {et~al.}(2009)\citenamefont {Eble},
  \citenamefont {Testelin}, \citenamefont {Desfonds}, \citenamefont
  {Bernardot}, \citenamefont {Balocchi}, \citenamefont {Amand}, \citenamefont
  {Miard}, \citenamefont {Lema{\^\i}tre}, \citenamefont {Marie},\ and\
  \citenamefont {Chamarro}}]{Eble2009}%
  \BibitemOpen
  \bibfield  {author} {\bibinfo {author} {\bibfnamefont {B.}~\bibnamefont
  {Eble}}, \bibinfo {author} {\bibfnamefont {C.}~\bibnamefont {Testelin}},
  \bibinfo {author} {\bibfnamefont {P.}~\bibnamefont {Desfonds}}, \bibinfo
  {author} {\bibfnamefont {F.}~\bibnamefont {Bernardot}}, \bibinfo {author}
  {\bibfnamefont {A.}~\bibnamefont {Balocchi}}, \bibinfo {author}
  {\bibfnamefont {T.}~\bibnamefont {Amand}}, \bibinfo {author} {\bibfnamefont
  {A.}~\bibnamefont {Miard}}, \bibinfo {author} {\bibfnamefont
  {A.}~\bibnamefont {Lema{\^\i}tre}}, \bibinfo {author} {\bibfnamefont
  {X.}~\bibnamefont {Marie}}, \ and\ \bibinfo {author} {\bibfnamefont
  {M.}~\bibnamefont {Chamarro}},\ }\bibfield  {title} {\enquote {\bibinfo
  {title} {{Hole{\textendash}Nuclear Spin Interaction in Quantum Dots}},}\
  }\href@noop {} {\bibfield  {journal} {\bibinfo  {journal} {Phys. Rev. Lett.}\
  }\textbf {\bibinfo {volume} {102}},\ \bibinfo {pages} {146601} (\bibinfo
  {year} {2009})}\BibitemShut {NoStop}%
\bibitem [{\citenamefont {Testelin}\ \emph {et~al.}(2009)\citenamefont
  {Testelin}, \citenamefont {Bernardot}, \citenamefont {Eble},\ and\
  \citenamefont {Chamarro}}]{Testelin2009}%
  \BibitemOpen
  \bibfield  {author} {\bibinfo {author} {\bibfnamefont {C.}~\bibnamefont
  {Testelin}}, \bibinfo {author} {\bibfnamefont {F.}~\bibnamefont {Bernardot}},
  \bibinfo {author} {\bibfnamefont {B.}~\bibnamefont {Eble}}, \ and\ \bibinfo
  {author} {\bibfnamefont {M.}~\bibnamefont {Chamarro}},\ }\bibfield  {title}
  {\enquote {\bibinfo {title} {{Hole{\textendash}spin dephasing time associated
  with hyperfine interaction in quantum dots}},}\ }\href@noop {} {\bibfield
  {journal} {\bibinfo  {journal} {Phys. Rev. B}\ }\textbf {\bibinfo {volume}
  {79}},\ \bibinfo {pages} {195440} (\bibinfo {year} {2009})}\BibitemShut
  {NoStop}%
\bibitem [{\citenamefont {Fallahi}\ \emph {et~al.}(2010)\citenamefont
  {Fallahi}, \citenamefont {Y{\i}lmaz},\ and\ \citenamefont
  {Imamoglu}}]{Fallahi2010}%
  \BibitemOpen
  \bibfield  {author} {\bibinfo {author} {\bibfnamefont {P}~\bibnamefont
  {Fallahi}}, \bibinfo {author} {\bibfnamefont {S~T}\ \bibnamefont
  {Y{\i}lmaz}}, \ and\ \bibinfo {author} {\bibfnamefont {A}~\bibnamefont
  {Imamoglu}},\ }\bibfield  {title} {\enquote {\bibinfo {title} {{Measurement
  of a Heavy-Hole Hyperfine Interaction in InGaAs Quantum Dots Using Resonance
  Fluorescence}},}\ }\href@noop {} {\bibfield  {journal} {\bibinfo  {journal}
  {Phys. Rev. Lett.}\ }\textbf {\bibinfo {volume} {105}},\ \bibinfo {pages}
  {257402} (\bibinfo {year} {2010})}\BibitemShut {NoStop}%
\bibitem [{\citenamefont {Sokolov}\ \emph {et~al.}(2017)\citenamefont
  {Sokolov}, \citenamefont {Petrov}, \citenamefont {Kavokin}, \citenamefont
  {Kurdyubov}, \citenamefont {Kuznetsova}, \citenamefont {Cherbunin},
  \citenamefont {Verbin}, \citenamefont {Poletaev}, \citenamefont {Yakovlev},
  \citenamefont {Suter},\ and\ \citenamefont {Bayer}}]{Sokolov2017}%
  \BibitemOpen
  \bibfield  {author} {\bibinfo {author} {\bibfnamefont {P.~S.}\ \bibnamefont
  {Sokolov}}, \bibinfo {author} {\bibfnamefont {M.~Yu.}\ \bibnamefont
  {Petrov}}, \bibinfo {author} {\bibfnamefont {K.~V.}\ \bibnamefont {Kavokin}},
  \bibinfo {author} {\bibfnamefont {A.~S.}\ \bibnamefont {Kurdyubov}}, \bibinfo
  {author} {\bibfnamefont {M.~S.}\ \bibnamefont {Kuznetsova}}, \bibinfo
  {author} {\bibfnamefont {R.~V.}\ \bibnamefont {Cherbunin}}, \bibinfo {author}
  {\bibfnamefont {S.~Yu.}\ \bibnamefont {Verbin}}, \bibinfo {author}
  {\bibfnamefont {N.~K.}\ \bibnamefont {Poletaev}}, \bibinfo {author}
  {\bibfnamefont {D.~R.}\ \bibnamefont {Yakovlev}}, \bibinfo {author}
  {\bibfnamefont {D.}~\bibnamefont {Suter}}, \ and\ \bibinfo {author}
  {\bibfnamefont {M.}~\bibnamefont {Bayer}},\ }\bibfield  {title} {\enquote
  {\bibinfo {title} {{Nuclear spin cooling by helicity-alternated optical
  pumping at weak magnetic fields in n-GaAs}},}\ }\href@noop {} {\bibfield
  {journal} {\bibinfo  {journal} {Phys. Rev. B}\ }\textbf {\bibinfo {volume}
  {96}},\ \bibinfo {pages} {205205} (\bibinfo {year} {2017})}\BibitemShut
  {NoStop}%
\bibitem [{\citenamefont {Paget}(1982)}]{Paget82}%
  \BibitemOpen
  \bibfield  {author} {\bibinfo {author} {\bibfnamefont {Daniel}\ \bibnamefont
  {Paget}},\ }\bibfield  {title} {\enquote {\bibinfo {title} {Optical detection
  of nmr in high-purity gaas: Direct study of the relaxation of nuclei close to
  shallow donors},}\ }\href {\doibase 10.1103/PhysRevB.25.4444} {\bibfield
  {journal} {\bibinfo  {journal} {Phys. Rev. B}\ }\textbf {\bibinfo {volume}
  {25}},\ \bibinfo {pages} {4444} (\bibinfo {year} {1982})}\BibitemShut
  {NoStop}%
\bibitem [{\citenamefont {Dyakonov}\ and\ \citenamefont
  {Perel}(1973)}]{DYAKONOV1973}%
  \BibitemOpen
  \bibfield  {author} {\bibinfo {author} {\bibfnamefont {M.~I.}\ \bibnamefont
  {Dyakonov}}\ and\ \bibinfo {author} {\bibfnamefont {V.~I.}\ \bibnamefont
  {Perel}},\ }\bibfield  {title} {\enquote {\bibinfo {title} {Optical
  orientation in a system of electrons and lattice nuclei in semiconductors -
  theory},}\ }\href@noop {} {\bibfield  {journal} {\bibinfo  {journal} {JETP}\
  }\textbf {\bibinfo {volume} {65}},\ \bibinfo {pages} {362} (\bibinfo {year}
  {1973})}\BibitemShut {NoStop}%
\bibitem [{\citenamefont {Paget}\ \emph {et~al.}(2008)\citenamefont {Paget},
  \citenamefont {Amand},\ and\ \citenamefont {Korb}}]{PagetAmand}%
  \BibitemOpen
  \bibfield  {author} {\bibinfo {author} {\bibfnamefont {D.}~\bibnamefont
  {Paget}}, \bibinfo {author} {\bibfnamefont {T.}~\bibnamefont {Amand}}, \ and\
  \bibinfo {author} {\bibfnamefont {J.~P.}\ \bibnamefont {Korb}},\ }\bibfield
  {title} {\enquote {\bibinfo {title} {{Light-induced nuclear quadrupolar
  relaxation in semiconductors}},}\ }\href@noop {} {\bibfield  {journal}
  {\bibinfo  {journal} {Phys. Rev. B}\ }\textbf {\bibinfo {volume} {77}},\
  \bibinfo {pages} {245201} (\bibinfo {year} {2008})}\BibitemShut {NoStop}%
\bibitem [{\citenamefont {Brun}\ \emph {et~al.}(1963)\citenamefont {Brun},
  \citenamefont {Mahler}, \citenamefont {Mahon},\ and\ \citenamefont
  {Pierce}}]{Brun}%
  \BibitemOpen
  \bibfield  {author} {\bibinfo {author} {\bibfnamefont {E.}~\bibnamefont
  {Brun}}, \bibinfo {author} {\bibfnamefont {R.~J.}\ \bibnamefont {Mahler}},
  \bibinfo {author} {\bibfnamefont {H.}~\bibnamefont {Mahon}}, \ and\ \bibinfo
  {author} {\bibfnamefont {W.~L.}\ \bibnamefont {Pierce}},\ }\bibfield  {title}
  {\enquote {\bibinfo {title} {{Electrically {I}nduced {N}uclear {Q}uadrupole
  {S}pin {T}ransitions in a {G}a{A}s {S}ingle {C}rystal}},}\ }\href@noop {}
  {\bibfield  {journal} {\bibinfo  {journal} {Phys. Rev.}\ }\textbf {\bibinfo
  {volume} {129}},\ \bibinfo {pages} {1965} (\bibinfo {year}
  {1963})}\BibitemShut {NoStop}%
\bibitem [{\citenamefont {Shklovskii}\ and\ \citenamefont
  {Efros}(1984)}]{Shklovskii&EfrosCh3}%
  \BibitemOpen
  \bibfield  {author} {\bibinfo {author} {\bibfnamefont {B.I.}\ \bibnamefont
  {Shklovskii}}\ and\ \bibinfo {author} {\bibfnamefont {A.L.}\ \bibnamefont
  {Efros}},\ }\href {http://books.google.ru/books?id=-0YsAAAAYAAJ} {\emph
  {\bibinfo {title} {Electronic Properties of Doped Semiconductors}}},\
  Springer series in solid-state sciences\ (\bibinfo  {publisher}
  {Springer-Verlag, Berlin},\ \bibinfo {year} {1984})\ Chap.~\bibinfo {chapter}
  {3}\BibitemShut {NoStop}%
\bibitem [{\citenamefont {Flisinski}\ \emph {et~al.}(2010)\citenamefont
  {Flisinski}, \citenamefont {Gerlovin}, \citenamefont {Ignatiev},
  \citenamefont {Petrov}, \citenamefont {Verbin}, \citenamefont {Yakovlev},
  \citenamefont {Reuter}, \citenamefont {Wieck},\ and\ \citenamefont
  {Bayer}}]{Petrov}%
  \BibitemOpen
  \bibfield  {author} {\bibinfo {author} {\bibfnamefont {K.}~\bibnamefont
  {Flisinski}}, \bibinfo {author} {\bibfnamefont {I.~Ya.}\ \bibnamefont
  {Gerlovin}}, \bibinfo {author} {\bibfnamefont {I.~V.}\ \bibnamefont
  {Ignatiev}}, \bibinfo {author} {\bibfnamefont {M.~Yu.}\ \bibnamefont
  {Petrov}}, \bibinfo {author} {\bibfnamefont {S.~Yu.}\ \bibnamefont {Verbin}},
  \bibinfo {author} {\bibfnamefont {D.~R.}\ \bibnamefont {Yakovlev}}, \bibinfo
  {author} {\bibfnamefont {D.}~\bibnamefont {Reuter}}, \bibinfo {author}
  {\bibfnamefont {A.~D.}\ \bibnamefont {Wieck}}, \ and\ \bibinfo {author}
  {\bibfnamefont {M.}~\bibnamefont {Bayer}},\ }\bibfield  {title} {\enquote
  {\bibinfo {title} {{Optically detected magnetic resonance at the
  quadrupole-split nuclear states in (In,Ga)As/GaAs quantum dots}},}\
  }\href@noop {} {\bibfield  {journal} {\bibinfo  {journal} {Phys. Rev. B}\
  }\textbf {\bibinfo {volume} {82}},\ \bibinfo {pages} {081308} (\bibinfo
  {year} {2010})}\BibitemShut {NoStop}%
\bibitem [{\citenamefont {Abragam}(1961)}]{Abragam}%
  \BibitemOpen
  \bibfield  {author} {\bibinfo {author} {\bibfnamefont {A.}~\bibnamefont
  {Abragam}},\ }\href@noop {} {\emph {\bibinfo {title} {The Principles of
  Nuclear Magnetism}}}\ (\bibinfo  {publisher} {Oxford University Press,
  Oxford},\ \bibinfo {year} {1961})\BibitemShut {NoStop}%
\end{thebibliography}

%


\end{document}